\documentstyle[12pt]{article}
\title{\bf The dependence of the nuclear charge form factor on short
range correlations and surface fluctuation effects}
\author{ S.E.Massen, V.P.Garistov$^+$ and M.E.Grypeos\\
Department of Theoretical Physics\\
Aristotle University of Thessaloniki\\
 GR--54006 Thesssaloniki, GREECE\\
 +Also:Institute of Nuclear Research and Nuclear Energy,\\
Bulgarian Academy of Sciences, Sofia, Bulgaria}

\begin{document}

\maketitle
\date{}

\begin{abstract}
We investigate the effects of fluctuations of the nuclear surface on the
harmonic oscillator elastic charge form factor of light nuclei,
while simultaneously approximating the short-range correlations
through a Jastrow correlation ~factor.
Inclusion of surface-fluctuation effects within this description, by
truncating
the cluster expansion at the two-body part, is found to improve somewhat the
fit to the elastic charge form-factor of $^{16}O$ and $^{40}Ca$.
However, the convergence of the cluster expansion is expected to deteriorate.
An additional finding is that the surface-fluctuation correlations produce a
drastic change in the asymptotic behavior of the point-proton form factor,
which now falls off quite slowly (i.e. as $const. \cdot q^{-4}$) at large
values of the momentum transfer $q$.

\end{abstract}

\newpage

\section{Introduction}
The calculation of the charge form factors $F_{ch}(q)$ of nuclei is a
challenging and appealing problem \cite{Elton67}. A possibility to face this
problem is by means of an independent-particle model.
In this approach, which is particularly attractive because of its simplicity,
the choice of the single particle potential has to be suitably made.
In fact a short range repulsion in this potential seems advisable for light
nuclei (see ref.\cite{Grypeos89} and references therein).
For example, with an harmonic oscillator (HO) potential having in addition
an infinite soft core of the form $\frac{B}{r^2}$ ($B>0$) the $F_{ch}(q)$ of
$^4He$ can be well reproduced, but for the heavier nuclei, such as $^{12}C$
and $^{16}O$, state dependent potentials seem necessary and even then the
fit is not so good for higher q-values \cite{Grypeos89}. Furthermore, the
correction of the centre of mass motion can not be made exactly and
unambiguously. An approach which is rather similar is the approach of Ripka
and Gillespie \cite{Ripka70} and Gaudin et al \cite{Gaudin71}.
It was shown by these authors that if a Jastrow wave function consisting
of HO orbitals and of simple state-independent correlation functions of the
form $f(r)=( 1-e^{-\beta ^2 r^2} )^{1/2}$ is used, one can
construct a Slater determinant which yields a density "very similar" to that
of such a Jastrow wave function in the two-body approximation. This is done
by diagonalizing the density matrix:
$$
\rho _{ij} = <\Psi |a^+_ja_i| \Psi>
$$
that is by using the so-called "natural orbitals". It is clear, however, that
the actual density matrix is unknown and one obtains it approximately,
usually by means of a Jastrow wave function, as is the case in the
above references (see expression (7.7) of the paper by Gaudin et al
\cite{Gaudin71}). Therefore, even with such an approach one still has in
practice to use a complicated wave function such as a Jastrow one.
{}From the above discussion it is clear that a Jastrow wave function is useful
even if one is only interested in calculating the form factor and density
distribution of nuclei and not necessarily other quantities, such as the
momentum distribution \cite{Bohigas80}.

In a series of papers \cite{SEM88, SEM89, SEM90} an expression
of the elastic charge form factor, $F_{ch}(q)$, truncated at the two
body term, was derived using the factor
cluster expansion of Ristig et al \cite{Ristig-Clark, Clark79}.
This expression, which is a sum of one-body and two-body terms, depends on
the
harmonic oscillator (HO) parameter $b_{1}$ and the correlation parameter
$\lambda$ through a Jastrow type correlation function which introduces the
short range correlations (SRC).
The use of the $HO$ orbitals (as well as of the particular and
popular form of the Jastrow correlation function used) stems
mainly from the considerable simplification they imply. There
are, however, additional advantages, such as:
{\bf i.} The correction of the centre of mass motion can be done
exactly by means of the Tassie and Barker factor \cite{Tassie58}.
{\bf ii.} One can obtain analytically the asymptotic behaviour of
the form factor $F_{ch}(q)$.
In principle, of course, one should start with a Hartree-Fock
independent-particle model assuming an effective interaction and then
introduce correlations.
Such an approach is, however, computation-wise quite demanding. ~Although it
is more sophisticated and more satisfactory, it lacks the above mentioned two
advantages.

The fit of $F_{ch}(q)$ to the experimental data with the above mentioned
procedure was very good both for low and high values of momentum
transfer except for the values around the last maximum for $ ^{16}O $ and
$ ^{40}Ca $.
Better fit can be obtained if the parameter $\lambda$ is taken to be
state dependent but in this case there is a big number of parameters, six for
$^{16}O$ \cite{SEM88} and ten for $^{40}Ca$  \cite{Lal93}.

Another possible way to make the agreement between theory and
experiment better might be to introduce, in addition, other types of ground
state ("long range") correlations which have
been the subject of previous investigations by a number of authors (see, for
example, \cite{Gar76, Anton, Gar95, Reinhard79, Esbensen83, Barranco85,
Shlomo88}).
We focus our attention, as in ref.\cite{Esbensen83}, on fluctuations of
nuclear surface due to the zero point motion of collective surface vibrations
\cite{Bohr69, Rowe70} which can  affect the ground state charge density.
The presence of surface fluctuation correlations (SFC) introduce another
fitting parameter in addition to the HO and the SRC parameters.
Thus, it appears to be of interest to ~develop the relevant formalism and
to investigate what would be the effect, if any, of this additional parameter
to the best fit values of the other parameters and to the quality of the
fit. The aim of this paper is to report on results of certain
investigations towards this direction.

In section 2, the above SFC are introduced to the HO densities and analytic
expressions of the nuclear density, elastic form factor and of the n-th
moment of the density distribution are given for the nuclei $^4 He$, $^{16}O$
and $^{40}Ca$. In section 3 the introduction of the SFC in addition to
the SRC is studied. Numerical results are reported and discussed in section 4.

\section{ The effect of the surface fluctuations on the harmonic
 oscillator density and form factor }

Our starting point is the expression for the proton (or charge) density of a
nucleus which has been deformed through
the zero-point motions of the collective surface vibrations. This expression,
according to ref. \cite{Esbensen83} (see also ref. \cite{Gar76, Anton, Gar95}
for a rather similar expression) is the following:
\begin{equation}
\rho_{1\sigma}(r)= {1 \over{\sqrt{2\pi}\sigma}} \int_{-\infty}^{\infty}
\rho_{1}(r-\xi) \exp\left[-{(\xi-s_0)^2 \over 2\sigma^2} \right] d\xi
\label{den-lrc}
\end{equation}
where $\rho_{1}(r)$ is the uncorrelated density, $s_{0}$ is a correction
needed to conserve the number of particles in the correlated
ground state and $ \sigma$ is a measure of the effect of the zero point
fluctuations. The value of $\sigma$ is related to $\beta_{\lambda}$, the
deformation parameters for the states of multipolarity $ \lambda $ , with the
relation ${\sigma}^2 \simeq {R^{2} _{0} \over {4 \pi}}
                            \sum_{\lambda} \beta^{2}_{\lambda}(\tau =0)$
while the $\beta_{\lambda}$ parameters can be determined from the values of
$B(E_{\lambda})$ \cite{Esbensen83, Shlomo88}.

In (\ref{den-lrc}) we consider for $\rho_{1}(r)$  the HO proton density,
in which the centre of mass correction has been taken into account,
 for nuclei $^{4}He$ to $^{40}Ca$. This is given by the expression:
\begin{equation}
 \rho_{1}(r)= {1\over{Z \pi^{3/2}} } {1\over {\tilde {b}}_{1} ^{3}}
\exp{[-{r^2 \over {\tilde {b}}_{1}^{2} }] } \sum_{k=0}^{2} N_{2k}
({r\over {\tilde {b}}_{1} })^{2k}
\label{den-HO}
\end{equation}
where
\begin{eqnarray}
N_0&= & 2\eta _{1s}+6\left( 1-\frac{b_1^2}{\tilde{b}_1^2}\right) \eta
_{1p}+\left( 10-20\frac{b_1^2}{\tilde{b}_1^2}+10\frac{b_1^4}{\tilde{b
}_1^4}\right) \eta _{1d}+ \nonumber  \\
&& \left( 2-4\frac{b_1^2}{\tilde{b}_1^2}+5\frac{b_1^4}{\tilde{b}_1^4}
\right) \eta _{2s} \nonumber  \\
N_2&=&  4\eta _{1p}\frac{b_1^2}{\tilde{b}_1^2}+\left( \frac 83\frac{b_1^2
}{\tilde{b}_1^2}-\frac{20}3\frac{b_1^4}{\tilde{b}_1^4}\right) \eta
_{2s}+\left( \frac{40}3\frac{b_1^2}{\tilde{b}_1^2}-\frac{40}3\frac{b_1^4
}{\tilde{b}_1^4}\right) \eta _{1d} \nonumber  \\
N_4&= & \left( \frac 83\eta _{1d} + \frac 43\eta _{2s}\right) \frac{b_1^4}
{\tilde{b}_1^4}
\label{N-HO}
\end{eqnarray}
and $\tilde{b}_1^2=b_1^2\left( 1-\frac 1A\right) $, A is the mass number and
$ b_{1}=\sqrt{\hbar \over {m \omega} } $ the harmonic oscillator parameter.
$\eta_{nl}$ is the occupation probability ($0$ or $1$ in the present
treatment) of the $nl$ state.
It is easily checked that when $\tilde{b}_1 = b_1$, that is when the centre of
mass correction is not taken into account, the coefficients of the polynomial
in \ref{den-HO} are reduced to the well known expressions:
$$
N_{0}=2 \eta_{1s} + 3 \eta_{2s}\; ,\:
N_{2}=4 \eta_{1p} - 4 \eta_{2s}\; ,\;
N_{4}={8 \over 3} \eta_{1d} + {4 \over 3} \eta_{2s}
$$
General expressions of similar structure for the density and the form factor
in the HO model have been given in ref \cite{Kosmas92}.

{}From (\ref{den-lrc}) and (\ref{den-HO}) an analytic
expression of $\rho_{1\sigma}(r)$ can be derived. This is:
\begin{equation}
\rho_{1\sigma}(r)= {1 \over {Z \pi ^{3/2} }}
{1 \over{{\tilde {b}}_{1}^{2} \sqrt{{\tilde {b}}_{1}^{2} +2 \sigma^2}  }}
\exp {\left[-{(r-s_0)^2 \over{({\tilde {b}}_{1}^{2} +2\sigma^2)} }\right] }
\, \sum_{k=0}^{4} C_{k} r^{k}
\label{den(HO-lrc)}
\end{equation}
where the coefficients $C_k$ depend on $ N_0, \,N_2, \,N_4,\, \sigma,\, s_0$
and ${\tilde {b}}_1$ and are given by the following formulae:
\begin{eqnarray}
C_0& =&  N_0 + B^2\,
    \left( {{{ {\tilde {b}}_1}}^2}\,{\sigma ^2} + 2\,{\sigma ^4} +
      {{{ {\tilde {b}}_1}}^2}\,{{{ s_0}}^2} \right)\, N_2  + \nonumber\\
  &&  \left( 3\,{B^2}\,{\sigma ^4} +
      6\,{B^3}\,{{{ {\tilde {b}}_1}}^2}\,{\sigma ^2}\,{{{ s_0}}^2} +
      {B^4}\,{{{ {\tilde {b}}_1}}^4}\,{{{ s_0}}^4} \right) { N_4}\, \nonumber
 \\
C_1& =& -2\, B^2 \,{ {\tilde {b}}_1}^2  \, s_0 \,  N_2 -
   4\, B^4 \,{ {\tilde {b}}_1}^2 \,{ s_0}\,
    \left( 3\, { {\tilde {b}}_1}^2\, \sigma ^2 + 6\, \sigma ^4 +
      { {\tilde {b}}_1}^2 \, { s_0}^2 \right) \,{ N_4}  \nonumber \\
C_2& =& B^2\,{ {\tilde {b}}_1}^2\, N_2 +  6\, B^4 \, { {\tilde {b}}_1}^2\,
    \left( {{\tilde {b}}_1}^2 \, \sigma ^2 + 2\, \sigma ^4 +
      {{\tilde {b}}_1}^2\, {s_0}^2 \right) \,  N_4 \nonumber \\
C_3& =& -4\,B^4 \,{ {\tilde {b}}_1}^4 \, s_0 \,  N_4 \nonumber \\
C_4& =& {B^4}\,{{{ {\tilde {b}}_1}}^4}\,{ N_4}
\label{ck}
\end{eqnarray}
and  $B=1/({\tilde {b}}_1^2 +2 \sigma ^2)$

By using expression (\ref{den(HO-lrc)}) one can find an analytic expression
for the n{\it th} moment of the density. This is the following:
\begin{eqnarray}
<r^n>_{1\sigma} &=& {2\, {{\tilde {b}}_1}^n \over {Z \sqrt{\pi} } }
\exp[-s_0^2 \,B] \sum_{k=0}^{4} C_{k} {\tilde {b}}_1^k\,
\left( 1+{2 \sigma ^2 \over {\tilde {b}}_1 ^2} \right)
{}^{(k+n+2)/2} \times \nonumber \\
&  & \left[   \Gamma({k+n+3 \over 2})\, _{1}\!F_{1}({k+n+3 \over 2};\,
{1 \over 2};\, {s_0^2 \,B} ) \, \right. + \nonumber \\
& & \left. { 2\, s_0 \, \sqrt{B} } \, \Gamma({k+n+4 \over 2}) \,
 _{1}\!F_{1}({k+n+4 \over 2};\, {3 \over 2};\, {s_0^2 \,B})
  \right]
\label{r^n-exact}
\end{eqnarray}

An approximate expression for $<r^n>_{1\sigma}$ may be derived by truncation
of
the series at the second power for $\sigma$ and the first power for $s_0$ :
\begin{eqnarray}
<r^n>_{1\sigma} & \simeq & {2 {\tilde {b}}_{1}^n \over {Z \sqrt{\pi} } }
\sum_{k=0}^{4} C_{k} {\tilde {b}}_1^k \times \nonumber \\
 & & \left[  \left( 1+(k+n+2) {\sigma^2 \over {\tilde {b}}_1^2} \right)
\Gamma({k+n+3 \over 2}) \,  \right.  + \nonumber \\
& &\left. {2 s_0 \over {\tilde {b}}_1} \left( 1+(k+n+1){\sigma^2 \over
{\tilde {b}}_1^2} \right)
 \Gamma({k+n+4 \over 2})  \right]
\label{r^n-approx}
\end{eqnarray}

By taking into account that
$$ <r^{0}>_{1\sigma} = <r^{0}>_{1}=1 $$
the approximate expression for the parameter $s_{0}$ is:
\begin{equation}
s_{0}\simeq - {\sqrt{\pi}\over 4 } \; {\sigma^{2} \over {\tilde {b}}_1 } \;
{ 2 N_0 +N_{2} +{3 \over 2 }N_{4}  \over{N_0 +N_2 + 2 N_4 } }
\label{s0}
\end{equation}

That expression was used as a first approximation in our
calculations. More accurate values were obtained by varying $s_0$
until normalization of $\rho _{1\sigma}(r)$ was achieved to a good
approximation.
{}From expressions (\ref{r^n-approx}) and (\ref{s0}) and from the known
expression of the moments of the HO density one can find the approximate
expression of the contribution of the SFC, $\Delta <r^2>_{1\sigma}$, to the
mean square radius for nuclei ${}^4 He$ to ${}^{40} Ca$. This is given by the
following expression:
\begin{eqnarray}
\Delta <r^2>_{1\sigma} &\simeq &{2 \over Z} \sigma ^2 \left[
3(N_0\,+\,{3\over 2} N_2 \,+\, {15 \over 4}N_4) \,- \right.  \nonumber   \\
& & \left.  {(2N_0 \,+\,N_2 \,+\, {3\over 2} N_4)\,(N_0 \,+\,2 N_2 \,+\,
6 N_4) \over {N_0 \,+\, N_2 \,+\, 2N_4} } \right]
\label{Dr2}
\end{eqnarray}

Finally, for the elastic point proton form factor the well known expression in
Born approximation
\begin{equation}
F_{1\sigma}(q)=4\pi \int_{0}^{\infty} \rho _{1\sigma}(r)
{ \sin (qr) \over{qr}}r^{2}dr
\label{F.T.}
\end{equation}
is used. Substitution of $\rho _{1\sigma}(r)$ from (\ref{den(HO-lrc)}) leads
to
the following analytic expression of $F_{1\sigma}(q)$ in terms of the
confluent
hypergeometric function
\begin{equation}
F_{1\sigma}(q)={1 \over {Z \sqrt\pi} } \; {\sqrt{B} \over {\tilde {b}}_{1}^2 }
\; {1 \over q} \;  \sum_{k=0}^{4} C_{k} I_{k}
\label{f(HO-lrc1)}
\end{equation}
where
\begin{eqnarray}
I_{k}&=& {1 \over B^{(k+2)/2} } \exp[- {s_0}^2 \,B ] \times \nonumber \\
& &Im \, \left[ 2 \Gamma({k+2\over{2}}) \,
 {}_{1}\!F_{1}({k+2 \over 2 }\,; \, {1\over 2}\,; \, z^2 )  \, + \,
4 \Gamma({k+3\over 2 }) \,  z\,
{}_{1}\!F_{1}({k+3\over 2 }\,; \,{3\over 2}\,; \,z^2) \, \right] \nonumber \\
 &  &
\label{I(k1)}
\end{eqnarray}
The complex quantity $z$ is given by:
$z = \sqrt{B}s_0 +i {q /(2 \sqrt{B} )} \; $.

Expression (\ref{I(k1)} ) may be reduced to a somewhat more convenient form:
\begin{eqnarray}
F_{1\sigma}(q) & =& {1 \over Z}  \exp[-{q^2 \over 4B}] \;
  \sum_{n=0}^{2} \left[ {\tilde{C}}_{2n} \cos(qs_0) +
  {\tilde{\tilde{C}}}_{2n}
  {\sin(qs_0) \over q }\right] ({q\over 2 \sqrt{B}})^{2n} + \nonumber \\
& &{2 \over \sqrt{\pi} Z {\tilde {b}}_1^2 \sqrt{B} } \, \exp[- s_0^2 \,B]\,
{1\over q}\,
 Im [I]
\label{f(HO-lrc)}
\end{eqnarray}
where
\begin{equation}
I=\sum_{n=0}^2 {C_{2n} \over B^n} \Gamma (n+1)
      _{1}\!F_{1}(n+1; {1\over 2};z^2) +
  2z \, \sum_{n=0}^1 {C_{2n+1} \over B^{n+{1\over 2}} } \Gamma (n+2)
      _{1}\!F_{1}(n+2; {3\over 2};z^2 )
\label{I(k+1)}
\end{equation}
The coefficients
${\tilde{C}}_{2n}$ and ${\tilde{\tilde{C}}}_{2n}$ depend also on
$ N_0, \,N_2, \,N_4,\, \sigma,\, s_0$ and ${\tilde {b}}_1$ and are given by
the following
expressions:
\begin{eqnarray}
{\tilde{C}}_0 &=& N_0\, + \, {3\over 2} N_2\,\,+ {{15}\over 4} N_4 \, + \,
{ {\sigma ^2} \over { {\tilde {b}}_1}^2} \,(2 N_0\,+\, N_2 \,+{3\over 2} N_4 )
\nonumber \\
{\tilde{C}}_2 &=& - N_2\, -\,(5\,-\, 4\sigma ^2 B) N_4 \nonumber \\
{\tilde{C}}_4 &=&  {\tilde {b}}_1^2\,B\, N_4    \nonumber \\
{\tilde{\tilde{C}}}_0 &=& { s_0 \over  {\tilde {b}}_1^2 } \,
(2 N_0\, +\, N_2\,+ {3\over 2} N_4)   \nonumber \\
{\tilde{\tilde{C}}}_2 &=& -s_0\,B(2 N_2\,+\,6N_4)  \nonumber \\
{\tilde{\tilde{C}}}_4 &=& 2\,s_0 \, {\tilde {b}}_1^2\, B^2 \, N_4
\label{ckp}
\end{eqnarray}

It may be easily checked from expression (\ref{f(HO-lrc)}) that when the SFC
are switched off, that is when the limiting case
$\sigma \rightarrow 0$ is considered, expression (\ref{f(HO-lrc)}) for
$F_{1\sigma}(q)$ goes over to the well known harmonic oscillator one, as
should
be the case, on the basis of expressions (\ref{F.T.}) and (\ref{den-lrc}).
Furthermore, by using the asymptotic expansion of the confluent hypergeometric
function, we
find that the behaviour for $F_{1\sigma}(q)$ at large values of the momentum
transfer is the following:
\begin{equation}
F_{1\sigma}(q) \simeq {1 \over \sqrt{\pi} Z}\,
{ \exp[- s_0 ^2 \, B] \over {{\tilde {b}}_1^2 \sqrt{B} } }\,
\left( A_4 \,({q \over {2\sqrt{B}}})^{-4} \,+
\, A_6 \, ({q \over {2\sqrt{B}} })^{-6} \,+
\, A_8 \,({q \over {2\sqrt{B}} })^{-8} \right)
\label{f-asym}
\end{equation}
where
\begin{eqnarray}
A_4&=& - s_0\,C_0 -{1 \over {2\,B}}C_1  \nonumber \\
A_6&=& ( -3\,s_0 + 2\,B\,{s_0}^3 )C_0 +
        ( 3\,{s_0}^2 - {3 \over {2 B}}) C_1 + {3s_0\over B} C_2 +
\,{3 \over {2\,B^2}} C_3   \nonumber \\
A_8&=& (- {45 s_0 \over 4} + 15 B {s_0}^3 - 597 B^2 {s_0}^5 ) C_0
+  (-{45 \over {8 B}} + {{45 {s_0}^2} \over 2} +
{315 \over 2} B {s_0}^4 ) C_1 +  \nonumber \\
& & ( {{45\,s_0}\over {2\,B}} - 15\,{s_0}^3 ) C_2 +
   ( {45 \over {4 B^2}} - {{45 {s_0}^2} \over {2 B}} )C_3
-{{45 s_0}\over {2 B^2}} C_4
\label{coef-asy}
\end{eqnarray}
Thus, it is seen that for sufficiently large values of $q$, the form factor
tends to zero rather slowly, namely as the inverse fourth power of the
momentum transfer. On the contrary, the HO form factor goes rapidly to zero
for large $q$, namely as a Gaussian or as a Gaussian times an even power of
$q$ (depending on the nucleus).

The value of $q$ at which the $F_{1\sigma}(q)$ approaches the value given by
the asymptotic expression (\ref{f-asym})
does not seem to depend very strongly on the nucleus, at least
when the values of the parameters $b_1$ and $\sigma$ are determined in the way
described in the following two sections. In Fig. 1 the $F_{1\sigma}(q)$ has
been plotted for the ${}^{16} O$ nucleus,
using the values $b_1 =1.563 fm $ and $\sigma = 0.414 fm $ (see section 4),
together with its asymptotic behaviour $const. \cdot q^{-4}$ and the improved
asymptotic expression (\ref{f-asym}), respectively. It is seen that
$F_{1\sigma}(q)$ becomes close to the asymptotic behaviour $const.\cdot
q^{-4}$ at quite large values of the momentum transfer ( larger than
$10 fm^{-1}$ ) while essential convergence to the behavior (\ref{f-asym}) is
achieved at much smaller $q$ values. It might also be of interest to note
that people have assumed in the past \cite{Drecher74} a decrease of the form
factor of the type $Cq^{-4}$ in the region where no measurements
are performed: $q>q_{max}$ (measured), in order to obtain
error envelopes on the densities of nuclei. The present
analysis indicates that the inclusion of additional terms
of the type mentioned above seems appropriate in this sort of analysis.

There are two parameters in expression (\ref{f(HO-lrc)}), the HO parameter
$b_1$ and the SFC parameter $\sigma$ which can be determined from the
deformation parameters $\beta _{\lambda}$ associated with the low lying
collective states of the nucleus or can be treated for example as a free
parameter. In the latter case, the fit of the form factor
(\ref{f(HO-lrc)}) (after correcting it for the finite proton size
 \cite{SEM88}) to the experimental data
(refs. \cite{Frosh67, Arnold78} for ${}^4 He$, \cite{Sick70} for ${}^{16}O$
and \cite{Sinha73} for ${}^{40}Ca$ )
leads to zero value for the parameter $\sigma$ except for ${}^4 He$.
For ${}^4 He$, $\sigma$ is different from zero ($\sigma =0.706 fm$ and
$b_1=1.252 fm$) and the value of $\chi ^2$ is smaller compared to the one
obtained with the HO model. However, the diffraction
minimum is not reproduced. Because of these reasons the introduction of short
range correlations is advisable. This is done in the next section.


\section{The effect of the surface fluctuations and short range
correlations on the charge form factor and density}

A general expression for the charge form factor $F_{ch}(q)$ of light closed
shell nuclei was derived \cite{SEM88, SEM90} using the factor cluster
expansion of Ristig, Ter Low and Clark \cite{Ristig-Clark, Clark79}.
This formula was further simplified by using normalized correlated wave
functions of the relative motion which were ~parameterized through a Jastrow
type relative two-body wave function of the form:
\begin {equation}
\psi_{nlS}(r) = N_{nlS}[1-\exp(-\lambda r^{2}/b^{2})]
\phi_{nl}(r)
\label{Jastrow wf}
\end {equation}
where $N_{nlS}$ are the normalization factors, $\phi_{nl}(r)$
the harmonic oscillator wave functions and
$b=\sqrt{2} b_1$ is the HO parameter for the relative motion.
The expression for $F(q)$ is of the form
\begin{equation}
F(q)=F_{1}(q)  +F_{2}(q)          \label{f-src}
\end{equation}
$F_{1}(q)$ is the contribution of the one-body term to
$F(q)$:
\begin{equation}
 F_{1}(q)= {1\over Z } \exp{[-{b_{1}^{2} q^{2} \over 4 }] }
 \sum_{k=0}^{2} \tilde{N}_{2k} ({ {b_{1} q} \over 2 })^{2k}
\label{f-HO}
\end{equation}
where
\begin{equation}
\begin{array}{lll}
\tilde{N}_{0}=2(\eta _{1s} + \eta _{2s} + 3 \eta _{1p} + 5 \eta _{1d}) & , &
\tilde{N}_{2}=-{4 \over 3} (2 \eta _{2s} + 3 \eta _{1p} + 10 \eta _{1d})  \\
 & &  \\
 \tilde{N}_{4}= {1 \over 3} (4 \eta _{2s} +8 \eta _{1d}) & &
\end{array}
\label{f-HO-N}
\end{equation}
while $F_{2}(q)$ is the contribution of the two-body term to $F(q)$ and is a
function of $q^{2}$ through the matrix elements
$$
A^{{n}'{l}'{S}'}_{nlS}(j_{k}) = <\psi_{nlS}\vert j_{k}(qr/2) \vert
\psi_{{n}'{l}'{S'}}>
$$
It consists of simple polynomials and exponential functions of $q^2$.

The point proton density can be obtained from (\ref{f-src}) by Fourier
transforming $F(q)$. The density is also separated into two parts:
\begin{equation}
\rho(r)=\rho _{1}(r) +\rho _{2}(r)
\label{den-src}
\end{equation}
$\rho _{1}(r)$ and $\rho _{2}(r)$ are the Fourier transforms of $F_{1}(q)$
and
$F_{2}(q)$, ~respectively. $\rho _{1}(r)$ is given by expression
(\ref{den-HO})
(with ${\tilde b}_1 = b_1$) while $\rho _{2}(r)$ is calculated numerically
because of the complexity of $F_{2}(q)$ mainly for  $^{40}Ca$.

The correlation parameter $\lambda$ and the HO parameter $b_{1}$ were
determined by fitting $F_{ch}(q)=f_{p}(q)f_{CM}(q)F(q)$ to the experimental
charge form factor. $f_{p}(q) \: , \; f_{CM}(q)=exp[ \frac{b_1^2 q^2}{4A}]$
are the corrections due to
the finite proton size \cite{SEM88} and the centre of mass motion
\cite{Tassie58}, respectively.

As it was pointed out in the introduction, a possible way of
improving the quality of the fit in the approach outlined
previously, should be to take into account the
correlations originating from the fluctuation of the nuclear surface.
Thus in (\ref{den-src}), $\rho_{1}(r)$ is substituted by $\rho_{1\sigma}(r)$,
while $\rho _{2}(r)$ by $\rho_{2\sigma}(r)$, which results by using in
expression (\ref{den-lrc}) instead of $\rho_{1}(r)$ the Fourier transform
of $\tilde{F}_2(q)=f_{CM}(q)F_2(q)$. Therefore, instead of $\rho(r)$ given
by (\ref{den-src}) we have now $\rho_{\sigma}$:
\begin{equation}
\rho_{\sigma} (r)=\rho_{1\sigma}(r) +\rho_{2\sigma}(r)
\label{den-sfcsrc}
\end{equation}
The (total) point-proton form factor $F_{\sigma}(q)$ is then obtained
by summing the Fourier transforms of $\rho_{1\sigma}(r)$,
$F_{1\sigma}$ and of $\rho_{2\sigma},\: F_{2\sigma}$, that is:
\begin{equation}
F_{\sigma} =  F_{1\sigma} + F_{2\sigma}
\label{Ftot-sigma}
\end{equation}

It should be noted that because of the assumed form
of the correlation function the asymptotic behaviour of
$F_{\sigma}(q)$ for large $q$ is expected to be of the same functional
form as that of $F_{1\sigma}(q)$. The coefficients, however, in the
various even negative power terms of $q$ will be different.
Thus, the introduction of the SFC effects in the correlated
(through Jastrow correlations), H.O. wave function leads to
less steep decrease of $F_{\sigma}(q)$ at large $q$.
It might be of interest to point out that a slow ~decrease of the form factor
at large $q$ is also suggested from our numerical results.

Finally, the charge form factor is obtained by multiplying $F_{\sigma}(q)$
by the charge form factor of the proton, $f_p(q)$:
\begin{equation}
F_{ch}(q) = f_p(q) F_{\sigma}(q)
\label{Ftot-ch}
\end{equation}

Expression (\ref{Ftot-ch}) was used in fitting to the experimental values
of the charge form factor, by treating $b_1,\: \lambda$, and $\sigma$
as fitting parameters.

\section{Numerical results and discussion}

The best fit values of the three parameters in the form factor, as well as
the values of $\chi^{2}$, for the nuclei $^{16}O $ and
$ ^{40}Ca$ are displayed in {\bf table 1} where three cases are considered.
In {\bf case 1} there are no correlations of any kind while in {\bf case 2}
the SRC are included. These two cases have been studied in previous works
\cite{SEM88, SEM89, SEM90}. Finally, in {\bf case 3}, both the SFC and SRC
are included. In the case of $^4 He$ the best fit value of the parameter
$\sigma$ goes to zero in the least-squares fitting procedure. Therefore, the
present approach does not seem to be ~applicable to this nucleus.

The value $\sigma =0.293 \: fm $ for $^{40}Ca$ may be compared with the value
$\sigma =0.638 fm $ which is given in \cite{Barranco85}.
It is seen that the value of $\sigma$ is considerably smaller in the
present approach in comparison to that of ref. \cite{Barranco85}.
An analogous remark holds for the value of this parameter for $^{16}O$, if
{}~comparison is made with the value obtained in ref. \cite{Escudero87}.

It is further noted that the value of $\sigma$ for $^{16}O$ is larger than
that of $^{40}Ca$. This is in accordance with what one expects on the basis of
the variation of $\sigma^2$ with the mass number, which was studied in ref.
\cite{Escudero87} and is further elaborated in the Appendix of the present
paper. We also observe that if the parameter $k$ (see expression
(\ref{A-1})) is determined from our value of $\sigma$ for $^{16}O$,
the value of $\sigma$ we find for ${}^{40} Ca$ is not far from the value
predicted on the basis of expression (\ref{A-7}). It is
smaller than that value by less than $30\%$.

We observe also from table 1 that for ${}^{16}O$ and ${}^{40} Ca$, the
introduction of the SFC and the SRC decreases the values of parameters
$b_{1}$ and $\lambda$ while the value of $( b_{1}^{2} / \lambda )^{1/2} $
(which is the "actual correlation parameter", since small values of
$( b_{1}^{2} / \lambda )^{1/2} $ imply values of the correlation factors
closer to unity) is increased. That would deteriorate the convergence of the
cluster expansion.
This is indicated by the values of the so-called "healing" or "wound"
integrals (see ref. \cite{Brink67}) for the various states:
\begin{equation}
\eta_{nls}^2= \int _0^{\infty} \mid \psi_{nls}(r) - \phi_{nl}(r) \mid ^2 dr
\label{eta-1}
\end{equation}

It is expected that, the larger the value of $\eta^2$ the worse the
convergence of the cluster expansion of the density. With the correlated
relative, two-body wave
function (\ref{Jastrow wf}), the healing integral is given simply by
\begin{equation}
\eta_{nls}^2= 2 [1 + N_{nls} ( I_{nls} - 1 ) ]
\label{eta-2}
\end{equation}
where the normalization factors $N_{nls}$ and the integrals
$
I_{nls} = \int _0^{\infty} e^{-\lambda r^2 / b^2 } \phi_{nl}^2 dr
$
can be ~obtained analytically for the various states. For the lowest state,
its
{}~expression is fairly simple:
\begin{equation}
\eta_{00}^2= 2 \left\{ 1 - \left [1-(1+\lambda)^{-3/2}\right ]
\left [1-2(1+\lambda)^{-3/2} + (1+2 \lambda)^{-3/2} \right ] ^{-1/2} \right\}
\label{eta-3}
\end{equation}
and can be easily used for calculations. We have dropped the index $S$, since
we consider state-independent correlation functions. For ${}^{16}O$, in case 2
we find $\eta_{00}^2 =0.00717$ while in case 3, $\eta_{00}^2 =0.01379$,
indicating deterioration of the convergence of the cluster expansion.
Similar are the results for the higher states (e.g. for the $1p$ state we
have in case 2, $\eta_{01}^2 =0.00027$ and $\eta_{01}^2 =0.00083$ in case 3).
Analogous are the results for ${}^{40}Ca$. In this case, however, the increase
of the values of $\eta_{nl}^2$ is smaller in comparison with the
corresponding one for ${}^{16}O$, indicating milder deterioration of the
convergence of the cluster expansion.

It is clear regarding the results of case 3 in table 1 that the SFC effects
on both, the one-body and the two-body part of the cluster expansion of the
form factor have been taken into account. Quite a simpler approach would be
to take into account these effects only on the one-body part. Such an
approach, however, would not be appropriate since then both terms would not
be treated on the same footing. It might have been acceptable if the effect
of the SFC on the two-body part had been sufficiently small, which
unfortunately
does not appear to be the case. This is indicated by the change in the
best-fit values of the parameters. If the SFC effects are taken into account
only for the one-body part of the form factor, then the best fit values for
$^{16}O$ are: $b_1=1.647\: fm, \; \lambda=11.440, \; \sigma=0.224 \: fm $ and
for $^{40}Ca$: $b_1=1.814 \: fm, \; \lambda=11.786, \; \sigma=0.364 \: fm $.
It is seen that there is a noticable change in the values of these
parameters. Furthermore, the value of $\sigma$ for $^{16}O$ is smaller than
the value of $\sigma$ for $^{40}Ca$, which contradicts our expectations (see
Appendix).

For the two nuclei we have considered, the introduction of the SFC has the
effect of improving the fit of $F_{ch}(q)$ to the experimental data.
Although the introduction of SFC does not decrease the value of $\chi^2$
too much, its relative decrease for ${}^{16}O$ being larger than that of
${}^{40}Ca$, there is an improvement of the fit in the region of the large
$q$-values.
This can be seen in figures 2a and 3a  where the $F_{ch}(q)$ for $^{16}O$
and $^{40}Ca$ have been plotted with the best fit values of the parameters
and compared with the experimental $F_{ch}(q)$.
Another interesting ~observation one can make regarding these figures is the
slow decrease of $F_{ch}(q)$ at large $q$ in case 3 (in which the SRC and
SFC are included in the form factor), in comparison with cases 1 and 2.

It seems also appropriate to point out that the putative roles of the
mean-field, short range correlations and surface-fluctuation
effects get mixed up to some degree. Thus, (see table 1) the introduction
of SRC changes the value of the HO parameter and the additional introduction
of SFC not only does it change further this value but also the value of the
parameter determing the short range correlations. Furthermore, the SFC
effects which should predominantly influence the low and medium $q$ behaviour
of the form factor produce a substantial change in its values and its
functional behavior at
very large values of the momentum transfer. One could perhaps say that the
introduction of the SFC "relieves the burden" assumed by the SRC
(in the two-body approximation) in correcting the independent particle model,
so that they can better perform the function for which they were designed.

In figures 2b and 3b the (corresponding to $F_{ch}(q)$) charge densities
$\rho_{ch}(r)$ of
$^{16}O$ and $^{40}Ca$ have been plotted. In the same figures certain other
relevant quantities have also been plotted, ~namely:

{\bf $\alpha$.} The two-body part of the density $\rho_2(r)$
corrected for the finite proton size: $\rho_2^{\prime}(r)$ without SFC.
The dashed line $\rho_2^{\prime}(r)$ was obtained with the parameters of case
2 and the solid line $\rho_2^{\prime}(r)$ was obtained with the parameters of
case 3. It is seen that if the two-body part of the cluster expansion of the
charge density with the Jastrow correlation function and HO orbitals is
calculated with the parameters of case 3, its absolute values for various $r$
increase rather considerably (see solid line) in comparison with its absolute
values calculated with the parameters of case 2.

{\bf $\beta$.} The ~difference of the charge densities
$\Delta \rho _{ch}(r) = \rho _{ch}(HO+SRC+SFC) - \rho_{ch}(HO+SRC)$, that is
the difference between the (HO+SRC) charge densities with and without SFC,
calculated with the parameters of case 3. It is seen that
$\Delta\rho_{ ch}(r)$ is quite small and characterized by oscillations.

\vspace{5mm}
In summary, the present analysis suggests that the inclusion of the
correlations originating from the fluctuations of the nuclear surface in the
usual cluster expansion (truncated at the two-body part) of the charge form
factor of $^{16}O$ and $^{40}Ca$ leads to some improvement in the quality of
the fit to the experimental data. It is also found that the convergence of
the cluster expansion is expected to deteriorate. Furthermore, the inclusion
of these correlations has a drastic effect on the asymptotic
behaviour of the
point-proton form factor, which now falls off for large $q$ quite slowly,
that is as $const. \, q^{-4}$.

\vspace{1cm}
{\bf Acknowledgement.} We would like to thank Prof. F. Barranco for useful
correspondence. This work is part of a research program supported by the
General Secretariat of ~Research and Technology, through contract PENED
No 360/91.

\newpage
\setcounter{equation}{0}

\appendix

\section{Appendix}
\renewcommand{\theequation}{\Alph{section}.\arabic{equation}}

In this appendix we discuss the variation of the parameter
$\sigma ^2$ with the mass number $A$ on the basis of a phenomenological
analysis. We assume that $\sigma ^2$ is proportional
to the ratio: {\it (surface of the nucleus)/(volume of the nucleus)}.
This is suggested by the results and discussion of
ref. \cite{Escudero87}. Thus, we may approximately write:
\begin{equation}
\sigma ^2 = \frac{k}{R_{eq}}
\label{A-1}
\end{equation}
where $k$ is a proportionality constant and $R_{eq}$  the
equivalent uniform radius of the nucleus, defined by
\begin{equation}
<r^2>_{eq} = \frac{3}{5} R_{eq}^2 = <r^2>
\label{A-2}
\end{equation}

We consider that the nuclear density distribution is
approximated by a symmetrized Fermi function \cite{Lukyanov73}:
\begin{equation}
\rho_{SF}(r) = \rho_0 \left[
\left( 1+ \exp \left[ \frac{r-c}{a} \right] \right)^{-1} +
\left( 1+ \exp \left[ \frac{-r-c}{a} \right] \right)^{-1} \right]
\label{A-3}
\end{equation}
where $c$ is determined by the normalization condition and
given by \cite{Dask83, Grypeos91}:
\begin{eqnarray}
c&=&(\frac{1}{2})^{\frac{1}{3}} r_0 A^{\frac{1}{3}} \left\{
\left[1+ \left[1+ \frac{4}{27} \left( \frac{\pi a}{r_0 A^{\frac{1}{3}} }
\right)^6
\right] ^{\frac{1}{2}} \right ]^{\frac{1}{3}}   +
\left[1- \left[1+ \frac{4}{27} \left( \frac{\pi a}{r_0 A^{\frac{1}{3}} }
\right)^6
\right] ^{\frac{1}{2}} \right ]^{\frac{1}{3}} \right\} \nonumber \\
 &=& (\frac{1}{2})^{\frac{1}{3}} r_0 A^{\frac{1}{3}} \left[
1 - \frac{1}{3} \left( \frac{\pi a}{r_0 A^{\frac{1}{3}} } \right)^2 +
   \frac{1}{81} \left( \frac{\pi a}{r_0 A^{\frac{1}{3}} } \right)^6 +
  \frac{1}{243} \left( \frac{\pi a}{r_0 A^{\frac{1}{3}} } \right)^8 + \cdots
  \right]
\label{A-4}
\end{eqnarray}
The parameter $r_0$ is expressed in terms of $\rho_0$:
$r_0= \left ( \frac{3}{4 \pi \rho _0} \right )^{1/3}$

Thus, we may write:
\begin{equation}
<r^2> \simeq <r^2>_{SF} = \frac{3}{5} c^2
\left[ 1 + \frac{7}{3} \left( \frac{\pi a}{c} \right )^2 \right]
\label{A-5}
\end{equation}

On the basis of expressions (\ref{A-1}), (\ref{A-2}) and (\ref{A-5}),
we find:
\begin{equation}
\sigma^2 = \frac{k}{c}\left[1+ \left( \frac{\pi a}{c} \right)^2 \right]^{-1/2}
\label{A-6}
\end{equation}

Except for the very light nuclei, we may use only the
first terms of the expansion of the above expression in
powers of $A$ and we may therefore write:
\begin{equation}
\sigma^2 \simeq k r_0^{-1} A^{-1/3} \left[
1 - \frac{5}{6} \left( \frac{\pi a}{r_0 A^{1/3} } \right)^2 +
\frac{71}{72} \left( \frac{\pi a}{r_0 A^{1/3}} \right)^4  \right]
\label{A-7}
\end{equation}

The parameters $r_0$ and $a$ appearing in $c$ may be determined by a least
squares fit of  the RMS radius of the symmetrized Fermi distribution to
the experimental values of the RMS radius of nuclei: $<r^2>_{exp}^{1/2}$.
Considering the nuclei of table 2 of ref. \cite{Escudero87} and the
corresponding values of  $<r^2>^{1/2}_{exp}$ cited there, we find
$r_0 = 1.147 \: fm$ and $a=0.507 \:fm$. The quality of the
fit is very satisfactory.

The value of $k$ may also be determined by fitting to the values of
$\sigma ^2$ obtained in \cite{Escudero87}, shown in Table 2.
The corresponding best fit value is $k=2.028 \: fm^3$.

Our results for the RMS radii are displayed in table 2, where also the
results for $\sigma ^2$ obtained with (\ref{A-6}) and
(\ref{A-7}) are given, as well as those with the expression
$\sigma ^2 = k_1  A^{-\beta}$. The best fit values of the parameters in the
latter expression are $k_1=1.415 \: fm^3$ and $\beta=0.312$, which are
rather close to the corresponding values of ref. \cite{Escudero87}
($\sigma^2 = 1.64 A^{-0.35}$), shown also in table 2.

We finally observe that expression (\ref{A-7}) gives results very close to
those obtained with (\ref{A-6}). There is only a small difference for the
lighter elements.


\newpage
\begin{table}[tbh]
\caption {\sf The values of the parameters $b_1$, $\lambda$,
$(b_{1}^2/  \lambda)^{1/2}$ and $\sigma$, the $\chi ^2$ and the RMS radius
and the contributions to it from the HO density and the various correlations
for nuclei ${}^{16}O$ and ${}^{40}Ca$ found by fitting to the experimental
form  factors. Case 1 is referred to the HO form factor, Case 2 when the SRC
are included in the form factor and Case 3 when the SRC and SFC are included
in the form factor. All the quantities except $\lambda$ and $\chi^2$ are in
fm.} a: \cite{Sick70}, b: \cite{Sinha73}

\scriptsize
\begin{tabular}{c c c c c c c c r l c l}
\hline\hline
 & & &  & & &  & &              &           & &  \\
 & & &  & & &  & & $<r_{ch}^{2}$& $>^{1/2}$ & &  \\
\ Case &\ Nucleus&$b_{1}$&$\lambda$&$\sqrt{b_{1}^{2}\over {\lambda}}$
&$\sigma$& $\chi^{2}$ & & & & & \\
   &  &  &  &  &  &  &total& HO & SRC & SFC & Exper.\\
\hline
 & & & & & & & & & & &  \\
1&$^{16}O$ &1.786&      &     &     &9013 &2.728&2.728&     &     &   \\
2&$^{16}O$ &1.679&12.768&0.470&     &6226 &2.659&2.577&0.654&     &      \\
3&$^{16}O$ &1.563& 7.989&0.553&0.414&6002 &2.655&2.433&0.838&0.655&2.728$^a$\\
\\
1&$^{40}Ca$&1.950&      &     &     &26847&3.439&3.439&     &     &    \\
2&$^{40}Ca$&1.860&13.915&0.499&     &19930&3.420&3.289&0.936&     &      \\
3&$^{40}Ca$&1.849&10.356&0.575&0.293&19588&3.505&3.272&1.165&0.484&3.482$^b$\\
\hline
\end{tabular}
\end{table}
\bigskip
\begin{table}[tbh]
\caption {\sf The values of RMS radii (in $fm$) and of the parameter
$\sigma^2$ (in $fm^2$) for a number of nuclei using the results of
ref.[30] and the phenomenological expressions of the
Appendix}.

\scriptsize
\begin{tabular}{c c c c c c c c }
\hline\hline
  &  &  &  &  &  &  &  \\
A  & $<r^2>_{exp}^{1/2}$ & $<r^2>_{SF}^{1/2}$ & $\sigma^2$
Ref. \cite{Escudero87} &
$\sigma^2$Ref. \cite{Escudero87} & $\sigma^2$
& $\sigma^2$ & $\sigma^2$   \\
   &                 &                & (Table 1)&
$(1.64 A^{-0.35})$&$(1.415 A^{-0.312})$&expr.\ref{A-6}&expr.\ref{A-7}\\
\hline
    &      &      &       &       &       &       &      \\
16  & 2.73 & 2.76 & 0.576 & 0.621 & 0.597 & 0.570 & 0.588 \\
40  & 3.49 & 3.43 & 0.487 & 0.451 & 0.449 & 0.458 & 0.460 \\
90  & 4.26 & 4.29 & 0.360 & 0.340 & 0.348 & 0.366 & 0.367 \\
208 & 5.50 & 5.50 & 0.235 & 0.253 & 0.268 & 0.286 & 0.286 \\
\hline
\end{tabular}
\end{table}

\newpage
{\center {\bf Figure Caption} }

{\bf Fig. 1.} The elastic point form factor in the HO model with SFC:
$F_{1\sigma}(q)$ for ${}^{16} O$ with $b1=1.563 fm$ and $\sigma = 0.414 fm$
(solid line) and its asymptotic behaviour $const. q^{-4}$ (long dashed line)
together with the values of the asymptotic expression (\ref{f-asym})
(short dashed line).

\bigskip

{\bf Fig. 2.} The charge form factor (3a) and density distribution (3b)
of ${}^{16}O$.  The experimental points of the form factor are from ref.
\cite{Sick70}, while for the density from refs. \cite{Sick79, Malaguti82}.
(For various cases and notation see text ).

\bigskip

{\bf Fig. 3.} The charge form factor (4a) and density distribution (4b)
of ${}^{40}Ca$.  The experimental points of the form factor are from ref.
\cite{Sinha73}, while for the density from refs. \cite{Sick79, Malaguti82}.
(For various cases and notation see text ).


\newpage
\begin{thebibliography}{99}
\bibitem{Elton67}
\begin{description}
\item[\tt a.] L.R.B. Elton, {\it Nuclear Sizes} (Clarendon Press, Oxford
1961).
\item[\tt b.] H. \"{U}berall, {\it Electron Scattering from Complex Nuclei}
(Academic Press, New York and London 1971).

\item[\tt c.] R.C. Barrett and D.F. Jackson, {\it Nuclear Sizes and
Structure} (Clarendon Press, Oxford 1977)
\end{description}

\bibitem{Grypeos89}
\begin{description}
\item[\tt a.] M. Grypeos and K. Ypsilantis, J. Phys.
G: Nucl. Part. Phys. {\bf 15} (1989) 1397.
\item[\tt b.] K. Ypsilantis and M. Grypeos, 4th Hellenic Symposium on Nuclear
              Physics, Ioannina-Greece, (1993), 99 and J. Phys. G:Nucl. Part.
Phys. (in Press).
\end{description}

\bibitem {Ripka70} G. Ripka and J. Gillespie, Phys. Rev.
Let. {\bf 25} (1970) 1624.

\bibitem {Gaudin71} M. Gaudin, J. Gillespie and G. Ripka,
Nucl. Phys. {\bf A176} (1971) 237.

\bibitem {Bohigas80} O. Bohigas and S. Strigari, Phys. Lett. {\bf 95B}
(1980) 9.

\bibitem{SEM88} S.E. Massen, H.P. Nassena and C.P. Panos, J. Phys. G:
Nucl. Phys. {\bf 14} (1988) 753.

\bibitem{SEM89} S.E. Massen and C.P. Panos, J. Phys. G:
Nucl. Part. Phys. {\bf 15} (1989) 311.

\bibitem{SEM90} S.E. Massen, J. Phys. G: Nucl. Part. Phys., {\bf 16}
 (1990) 1713.

\bibitem{Ristig-Clark} M.L. Ristig, W.J. Ter Low and J.W. Clark, Phys. Rev.
{\bf C3}  (1971)  1504.

\bibitem{Clark79} J.W. Clark Prog. Part. Nucl. Phys. {\bf 2} (1979) 89.

\bibitem{Tassie58} L.J.Tassie and F.C.Barker, Phys. Rev. {\bf 111}
 (1958) 940.

\bibitem{Lal93} G.A. Lalazissis, S.E. Massen and C.P. Panos, Z. Physik
{\bf A348} (1994) 257.

\bibitem{Gar76} V.P.Garistov and I.J.Petkov, Bulg.J.Phys. {\bf 3} (1976) 6.

\bibitem{Anton} A.N. Antonov,V.P. Garistov and I.J. Petkov Phys. Let.{\bf 68B}
 (1977) 305.

\bibitem{Gar95} V.P.Garistov, Int. Journ. of Modern Physics E (in Press).

\bibitem{Reinhard79} P.G. Reinhard and D. Drechsel, Z. Physik {\bf A290}
 (1979) 85.

\bibitem{Esbensen83} H. Esbensen, and G.F.Bertsch Phys. Rev. {\bf C28}
 (1983) 355.

\bibitem{Barranco85} F. Barranco and R.A. Broglia Phys. Let. {\bf 151B}
 (1985) 90.

\bibitem{Shlomo88} S. Shlomo Phys. Lett. {\bf 209B} (1988) 23.

\bibitem{Bohr69} A. Bohr and B.R.Mottelson, {\it Nuclear Structure}
(Benjamin, New York, 1969).

\bibitem{Rowe70} D.J. Rowe, {\it Nuclear Collective Motion} (Methuen and
Co. TLD., London, 1970).

\bibitem{Kosmas92} T.S. Kosmas and J.D. Vergados Nucl. Phys. {\bf A536}
 (1992) 72.

\bibitem{Drecher74}
\begin{description}
\item[\tt a.] B. Drecher, J. Friedrich, K. Merle, H.
Rothhaas, and G. L\"{u}hrs, Nucl. Phys. A235 (1974) 219.
\item[\tt b.] H.J. Emrich, G. Frick, G. Mallot, H. Miska, and H.-G.
Sieberling, Nucl. Phys. {\bf A396} (1983) 401c.
\item[\tt c.]  R.Soundranayagam, A. Saha, K.K. Seth, C.W. deJager, H. deVries,
H. Blok, and G. van der Steenhoven, Phys. Lett. {\bf B212} (1988) 13.
\end{description}

\bibitem{Frosh67} R.F. Frosh, J.S. McCarthy and M.R. Yearian Phys. Rev.
{\bf 160} (1967) 874.

\bibitem{Arnold78} R.G. Arnold, B.T. Chertok, S. Rock, W.P. Schutz, S.M.
Szalata, D.Day, J.S. McCarthy, F.Martin, B.A.Mecking, I.Sick and G Tamas
Phys. Rev. Let. {\bf 40} (1978) 1429.

\bibitem{Sick70} I. Sick and J.S. McCarthy Nucl.Phys. {\bf A150} 631 (1970).

\bibitem{Sinha73} B.B. Sinha, G.A. Peterson, R.R. Withey, I. Sick and
J.C. McCarthy Phys. Rev. {\bf C7} (1973) 1930.

\bibitem{Sick79} I. Sick, Private communication with P. E. Hodgson and B. A.
Brown

\bibitem{Malaguti82} F. Malaguti, A. Uguzzoni, E. Verodini and P.E. Hodgson
Rivista Nuovo Cimento {\bf 5} (1982) 1.

\bibitem{Escudero87} J.I. Escudero, F. Barranco and G. Madurga
J. Phys. G: Nucl. Part. Phys., {\bf 13} (1987) 1261.

\bibitem{Brink67} D.M. Brink and M.E. Grypeos, Nucl. Phys. {\bf A97} (1967)
81.

\bibitem{Lukyanov73} Yu.N. Eldyshev, V.N. Lukyanov and Yu.S. Pol
Sov. J. Nucl. Phys. {\bf 16} (1973) 282.

\bibitem{Dask83} C. Daskaloyannis, M. Grypeos, C. Koutroulos, S Massen and
D. Saloupis, Phys. Lett. {\bf 121B} (1983) 91.

\bibitem{Grypeos91} M.E. Grypeos, G.A. Lalazissis, S.E. Massen and C.P.
Panos J. Phys. G: Nucl. Part. Phys., {\bf 17} (1991) 1093.
\end{thebibliography}
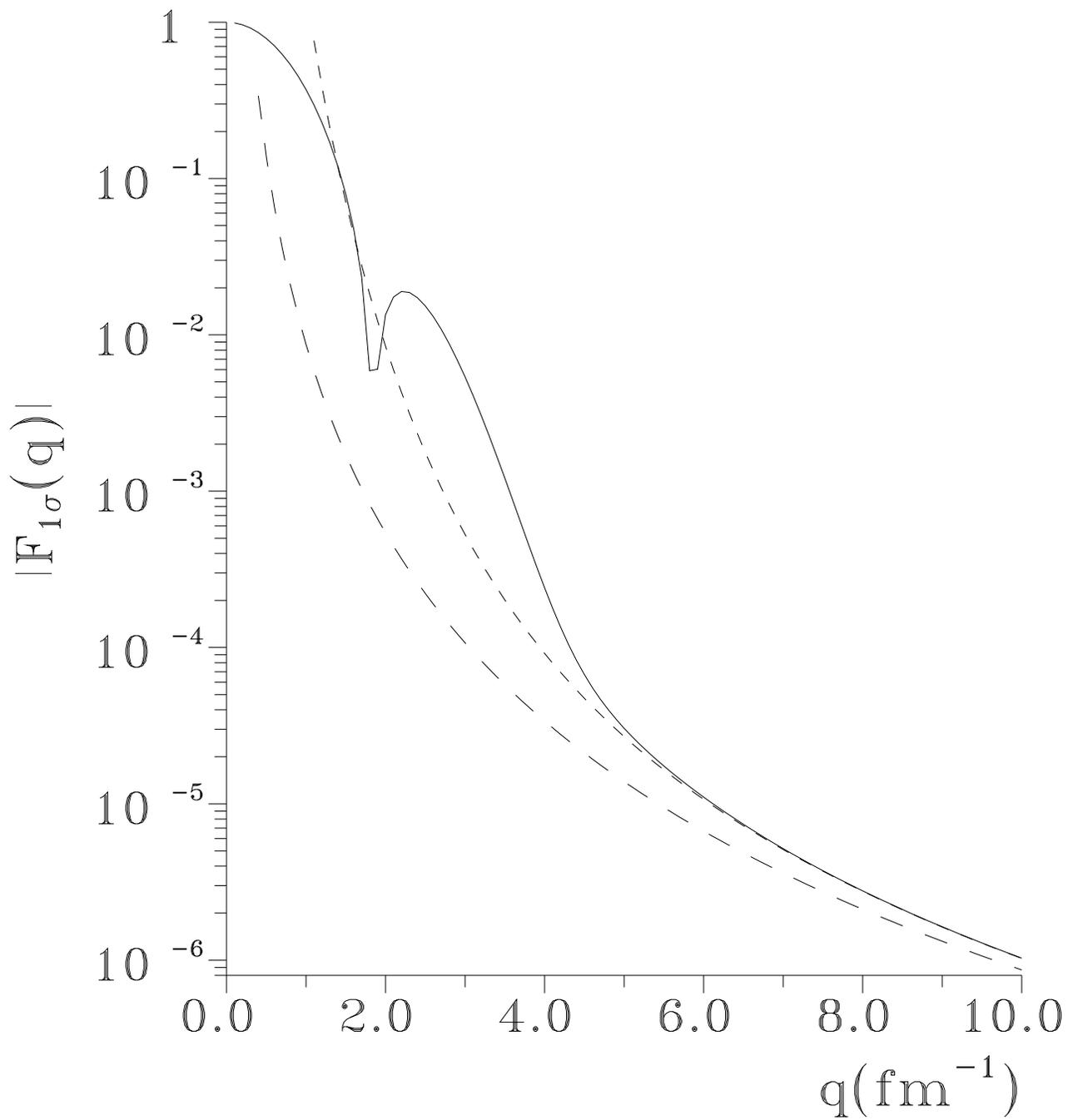
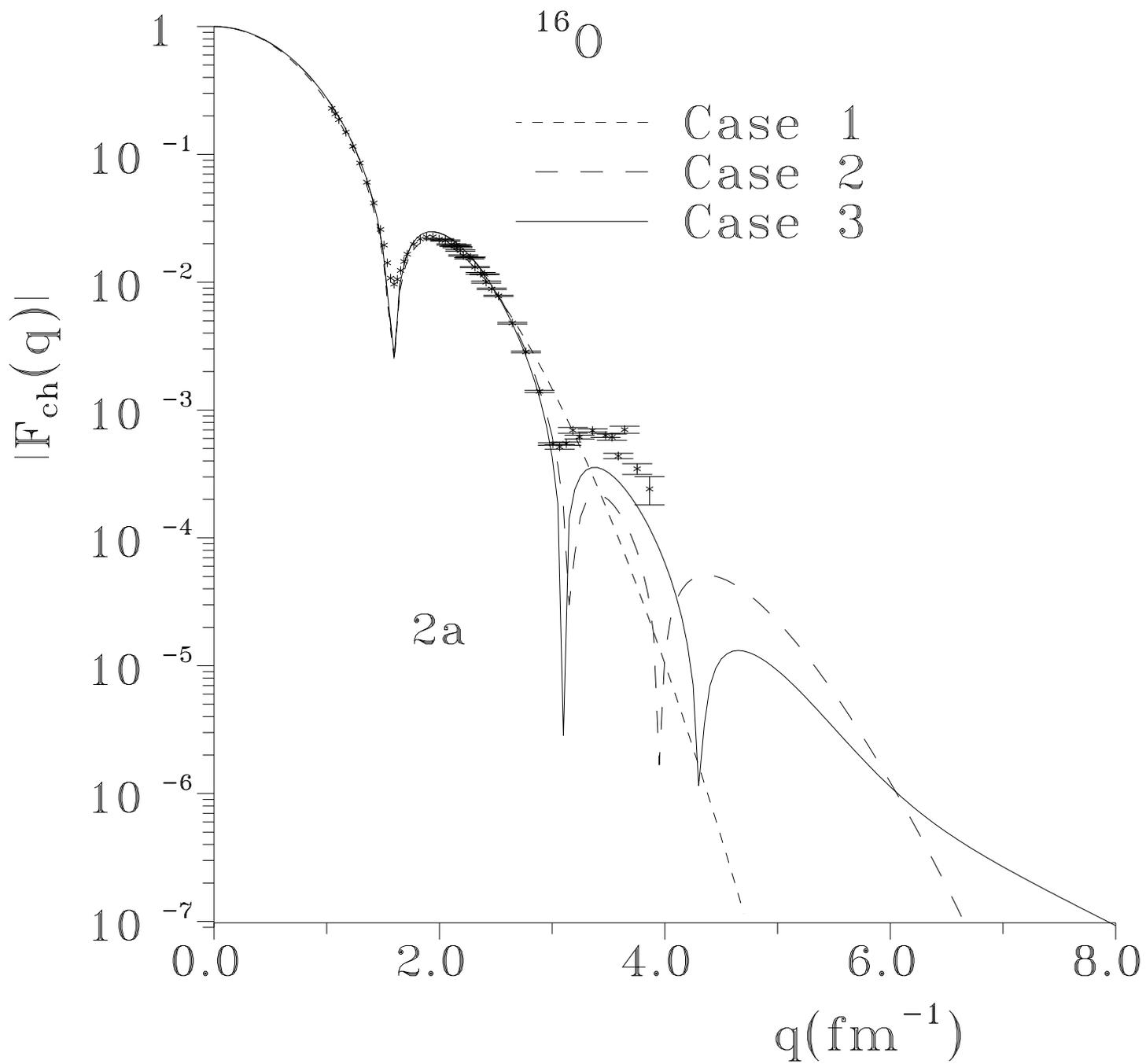
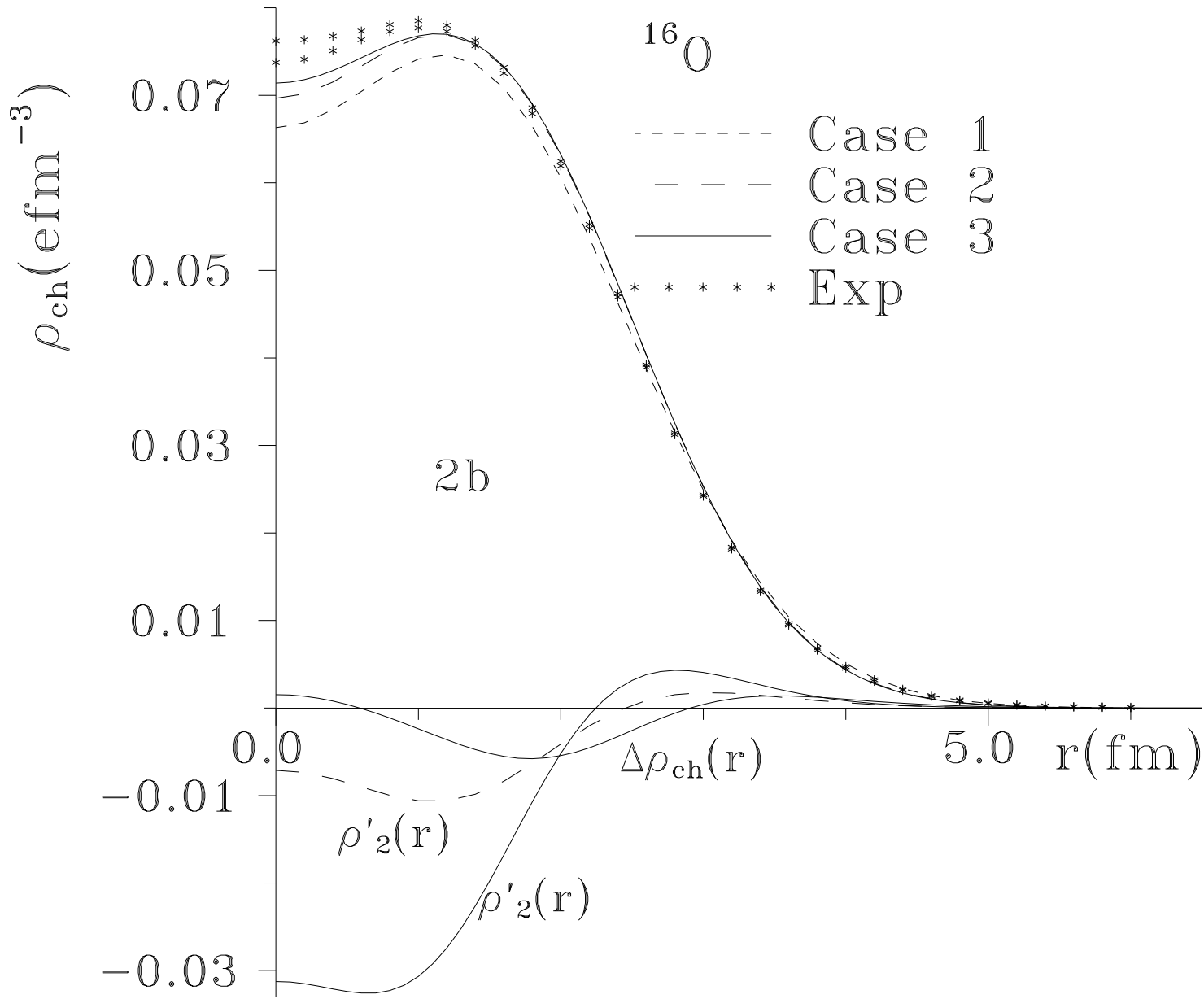
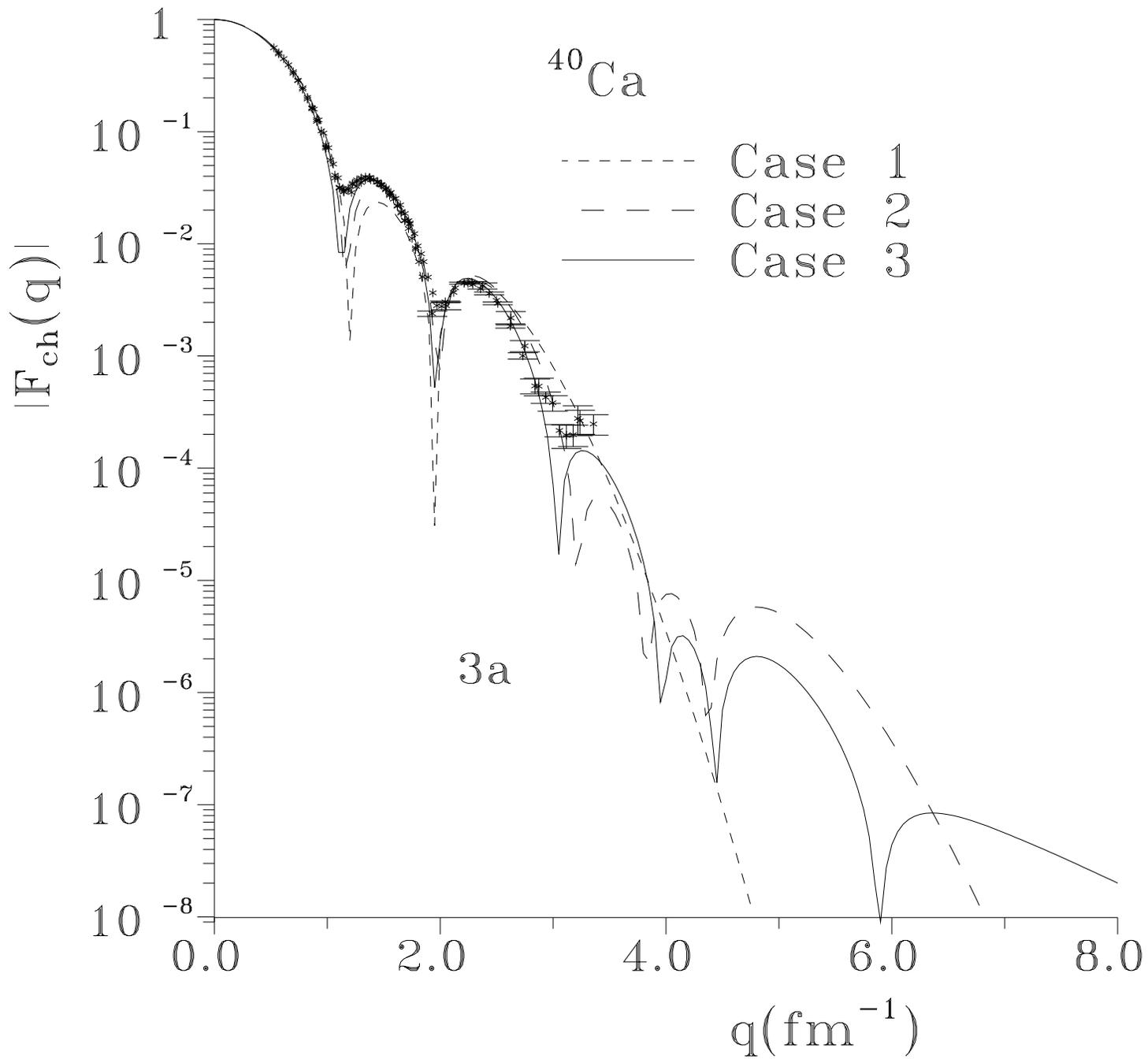
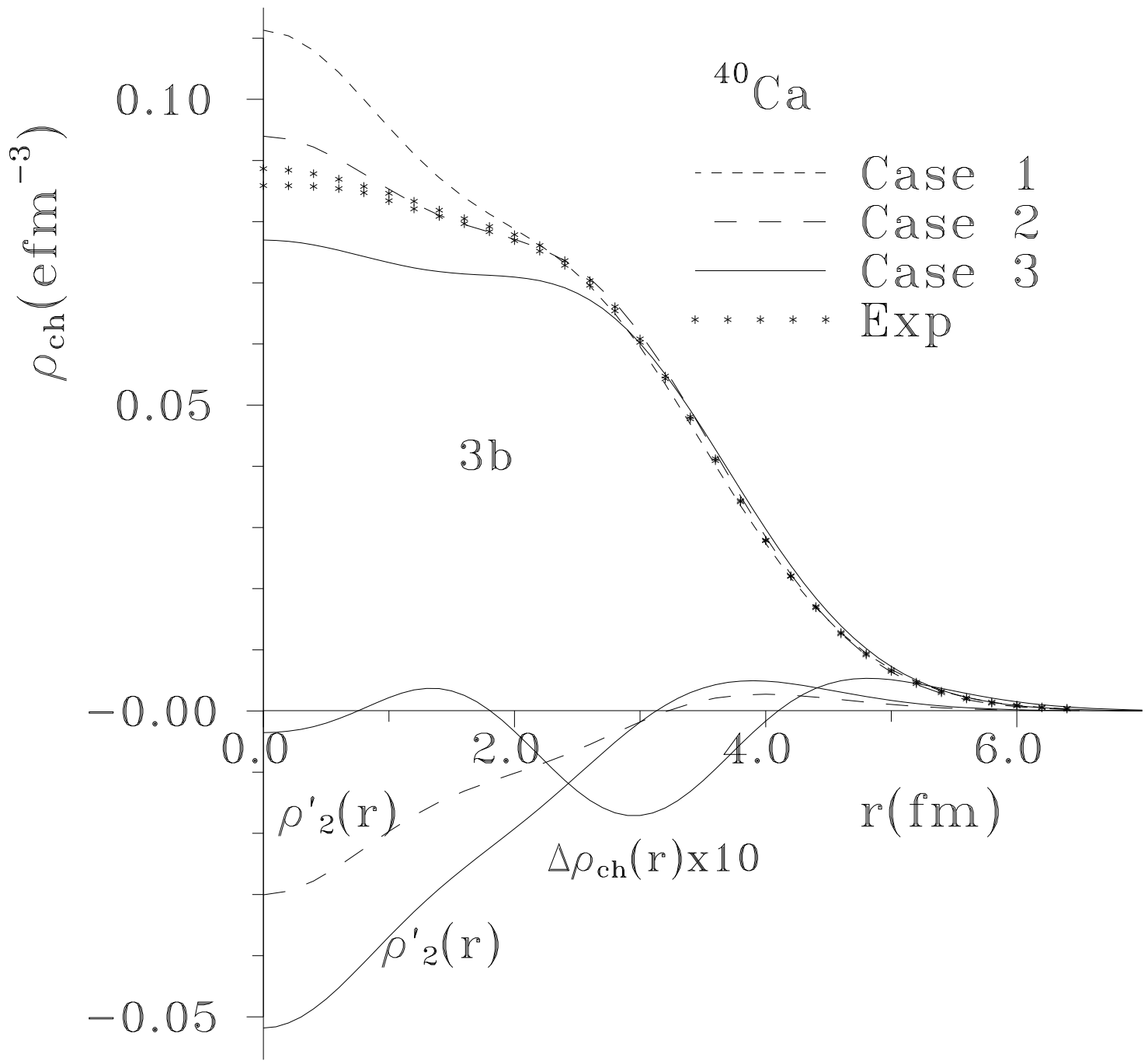
\end{document}